\documentclass[twocolumn,aps,prb,superscriptaddress,showpacs]{revtex4}

\usepackage{mathptmx}
\usepackage{subfigure}
\usepackage{dcolumn}
\usepackage{amsmath,amssymb}
\usepackage{bm}
\usepackage{color}
\usepackage{latexsym}
\usepackage[english]{babel}
\usepackage{times}

\usepackage{psfrag,graphicx} 
\usepackage{epsf}  
\usepackage{amsfonts}
\usepackage{natbib}
\usepackage{epstopdf}
\usepackage{appendix}
\usepackage[colorlinks=true,citecolor=blue,linkcolor=blue]{hyperref}

\def\bs#1{\boldsymbol{#1}}

\begin{document}


\title{Coexistence of spin-$1/2$ and spin-$1$ Dirac-Weyl fermions in the edge-centered honeycomb lattice}

\author{Zhihao Lan}
\affiliation{SUPA, Department of Physics, Heriot-Watt University, EH14 4AS, Edinburgh, United Kingdom}
\affiliation{School of Mathematics, University of Southampton, Highfield, Southampton, SO17 1BJ, United Kingdom}
\author{Nathan Goldman}
\affiliation{Center for Nonlinear Phenomena and Complex Systems - Universit\'e Libre de Bruxelles , 231, Campus Plaine, B-1050 Brussels, Belgium}

\author{Patrik \"Ohberg}
\affiliation{SUPA, Department of Physics, Heriot-Watt University, EH14 4AS, Edinburgh, United Kingdom}

\begin{abstract}

We investigate the properties of an edge-centered honeycomb lattice, and show that this lattice features both spin-$1/2$ and spin-$1$ Dirac-Weyl fermions at different filling fractions $f$  ( $f=1/5,4/5$ for spin $1/2$ and $f=1/2$ for spin $1$). This five-band system is the simplest lattice that can support simultaneously the two different paradigmatic Dirac-Weyl fermions with half-integer spin and integer spin. We demonstrate that these pseudo-relativistic structures, including a flat band at half-filling, can be deduced from the underlying Kagome sublattice.  We further show that the signatures of the two kinds of relativistic fermions can be clearly revealed by several perturbations, such as a uniform magnetic field, a Haldane-type spin-orbit term, and charge density waves. We comment on the possibility to probe the similarities and differences between the two kinds of relativistic fermions, or even to isolate them individually. We present a realistic scheme to realize such a system using cold atoms.

\end{abstract}

\maketitle 


\section{Introduction} 

In the solid state, matter is typically organised into crystal structures. The mathematical models for describing different materials are consequently based on lattices where the electrons are trapped in periodic structures. The understanding of the equilibrium and transport mechanisms in such systems also forms our knowledge of many fundamental effects and indeed technological applications of today. Recently emergent phenomena such as quasi-relativistic effects in non-relativistic settings have proven to be important in this respect. Most notably graphene \cite{graphene_rev}, together with topological insulators \cite{top_insulators_rev} and cold atoms in optical lattices \cite{moving_dirac_points,Lim2008,Goldman2009,Bermudez2010prl,Lee2009,na2}, are prominent examples of this.
The emergent quasi-relativistic fermions in graphene and topological insulators are two-spinor massless Dirac fermions, but it has also been suggested that in more exotic lattices, such as the $\mathcal{T}_3$ lattice ~\cite{weyl_1} and the line-centered-square Lieb lattice ~\cite{shen:2010, apaja:2010,goldmanlieb}, the emergent massless fermions are in fact  pseudospin-$1$ objects which also involve a flat band. Also Dirac-Weyl fermions with arbitrarily large spin have been studied based on fermionic atoms trapped in optical superlattices \cite{weyl_n,lattice_model}.

In this paper we explore a different direction: can we find a single setup where several different kinds of Dirac-Weyl fermions can coexist in the same lattice? This is an intriguing question. Such a  ``material" should have remarkably versatile 
properties as far as density dependent effects are concerned, as we will show in this paper. Our initial efforts to solve this problem is motivated by the study of Lieb lattices ~\cite{shen:2010, apaja:2010,goldmanlieb}, where additional lattice sites on the edges of the square lattice give rise to a flat band.  From this inspiration, we expect that when introducing additional lattice sites to the edges of the standard honeycomb lattice~\cite{graphene_rev}, which we refer to as the \emph{edge-centered honeycomb} (ECH) lattice in the following, a flat band should also emerge  \cite{DOS_T3}, thus giving both spin-$1/2$ and spin-$1$ Dirac-Weyl fermions. This is indeed what we have found in this study.  We also note that the next-nearest-neighbor (NNN) hopping in this ECH lattice produces the well-known Kagome lattice~\cite{kagome_TI}, thus the ECH lattice interpolates between several well researched lattices, such as honeycomb, Lieb and Kagome lattices.  In fact, by using an intriguing mapping (see Appendix~\ref{myappendix}), we find the band structure of the ECH lattice is completely determined by its underlying honeycomb and Kagome sublattices, revealing the deep connection between the ECH, the honeycomb, and the Kagome lattices. Furthermore, we investigate the response of the system to perturbations such as  a uniform magnetic field,  a Haldane-type spin-orbit coupling \cite{Kane_Mele}, and a charge density wave \cite{kagome_TI} (CDW). We demonstrate how these perturbations allow us to probe the similarities and differences between the two kinds of relativistic fermions. This five-band model turns out to be a minimal model that can support simultaneously the two different paradigmatic Dirac-Weyl fermions at the lowest spin level where  spin-$1/2$ Dirac points coexist with a single spin-$1$ Dirac point crossed by a flat band.



\section{The model and energy spectrum}

\begin{figure}[!hbp]
	\centering
	\includegraphics[width=1.\columnwidth]{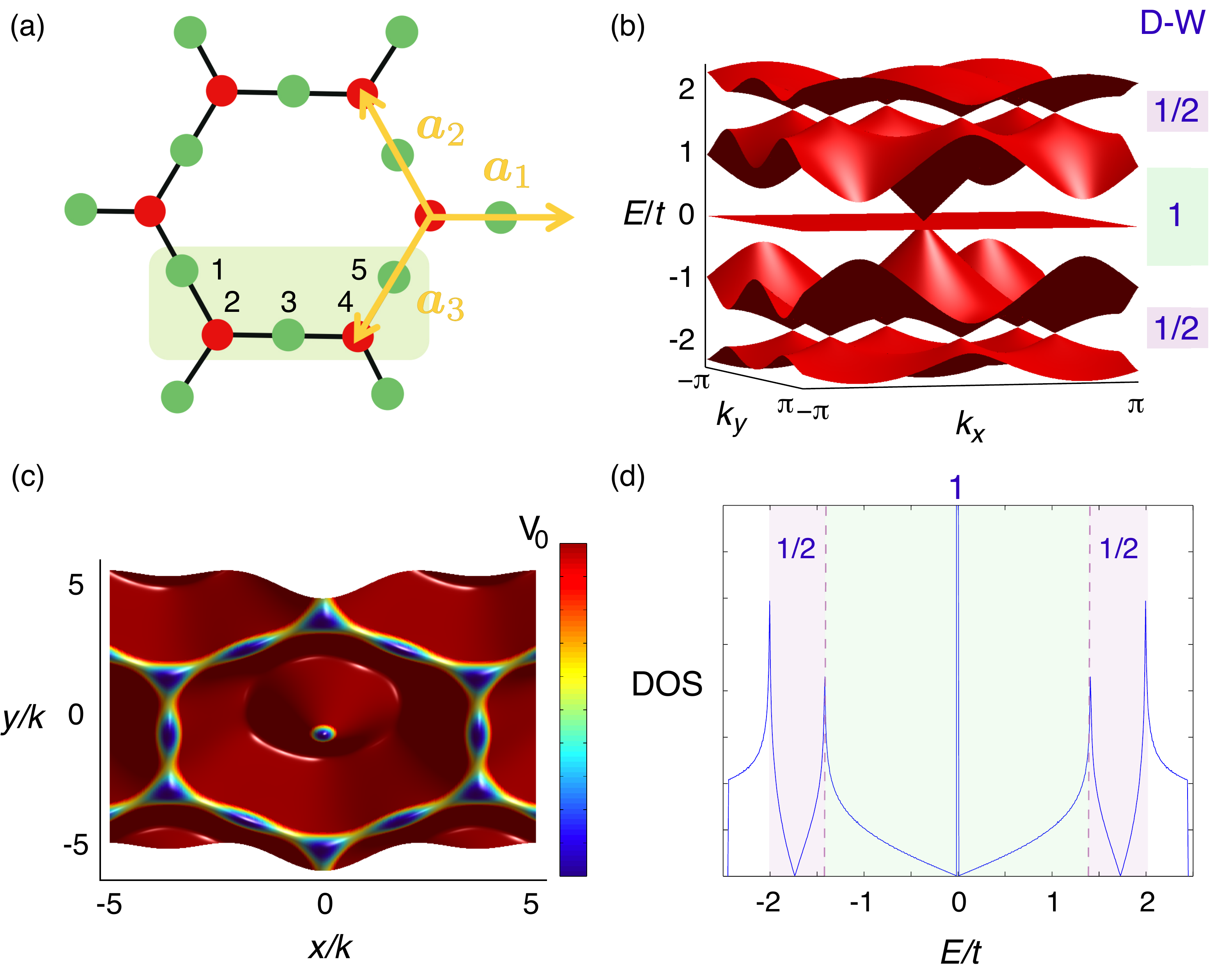}
	\caption{ (a) The ECH lattice is characterized by five sites per unit cell, denoted by $\tau=1, \dots, 5$. The three NN vectors of the underlying honeycomb structure (red sites) are given by ${\bf a}_1=(1,0)$, ${\bf a}_2=(-1/2,\sqrt{3}/2)$, ${\bf a}_3=(-1/2,-\sqrt{3}/2)$ while the NN vectors of the ECH lattice by ${\bf \nu}_{\mu}={\bf a}_{\mu}/2$, where $\mu=1,2,3$.  (b) The energy band structure $E(\bs{k})/t$ of the ECH lattice hosts two different kinds of Dirac-Weyl fermions: spin-1/2 at $f=1/5, 4/5$ and spin-1 at $f=1/2$.  (c) The intensity profile $\vert I (x,y) \vert$ obtained from the six-laser configuration in Eq. \eqref{fields} that would create an ECH lattice for cold atoms. (d). The density of states (DOS) of the band structure illustrated in (b). The two relativistic regimes (spin-$1/2$ and spin-$1$) are separated by van Hove singularities at $E = \pm \sqrt{2} t$ indicated by the vertical dotted lines.}
	\label{lattice}
\end{figure}

We are interested in the properties of a fermionic gas trapped in an ECH lattice whose unit cell contains 5 inequivalent sites, which are labeled by $\tau=1, \dots, 5$, as illustrated in Fig.  ~\ref{lattice} (a). A physical realization of this system could be achieved by trapping fermionic atoms using six lasers, which divide the plane into six sectors of $60^{\circ}$.  The corresponding configuration of the laser light fields can be chosen as 
\begin{align}
& {\bf E}_1=E(0,1)e^{i k {\bf x} \cdot {\bf a}_1}, & &{\bf E}_2=E(\sqrt{3}/2,1/2)e^{-i k {\bf x} \cdot {\bf a}_2}, \nonumber \\
&{\bf E}_3=E(\sqrt{3}/2,-/2)e^{i k {\bf x} \cdot {\bf a}_3}, & &{\bf E}_4=E(0,-1)e^{-i k {\bf x} \cdot {\bf a}_1},   \nonumber \\
&{\bf E}_5=E(-\sqrt{3}/2,-1/2)e^{i k {\bf x} \cdot {\bf a}_2}, & & {\bf E}_6=E(-\sqrt{3}/2,1/2)e^{-i k {\bf x} \cdot {\bf a}_3}, \label{fields}
\end{align} 
where ${\bf a}_1=(1,0)$, ${\bf a}_2=(-1/2,\sqrt{3}/2)$, ${\bf a}_3=(-1/2,-\sqrt{3}/2)$ are the three nearest-neighbor (NN) vectors of the underlying honeycomb structure (i.e., the red sites in Fig. ~\ref{lattice} (a)). In this case, the intensity profile $I(x,y)=\vert {\bf E}_{\text{tot}} (x,y) \vert ^2$  from the total electric field ${\bf E}_{\text{tot}}=\sum_i {\bf E}_i$ produces a potential landscape as shown in Fig. ~\ref{lattice} (c).  Alternatively, one can envisage using Spatial Light Modulators for shaping the intensity of a light beam such that the desired minima create the ECH lattice (see for instance Whyte and Courtial \cite{whyte2005} and references therein).  Also nanostructured lattice potentials for two-dimensional electron gases ~\cite{artificial_lattice} can be considered for creating the ECH lattice. 

For sufficiently deep lattice sites we can use the tight-binding approximation, which results in the non-interacting Hamiltonian 
\begin{equation}
H_0=t\sum_{\langle ij \rangle}c_i^{\dagger}c_j,\label{myham}
\end{equation}
where $c_i^{\dagger}$($c_i$) is the creation (annihilation) operator at the lattice site $i$ and $t$ is the nearest-neighbor (NN) hopping amplitude. The explicit single-particle Schr\"odinger equation derived from Eq. \eqref{myham} is detailed in Appendix \ref{myappendix}.
The energy band structure of this system can be obtained from the Hamiltonian in momentum space $H=\sum_{{\bf k}}\Psi^{\dagger}_{{\bf k}}H_{{\bf k}}\Psi_{{\bf k}}$, where $\Psi_{{\bf k}}=(c_{1{\bf k}}, c_{2{\bf k}},c_{3{\bf k}},c_{4{\bf k}},c_{5{\bf k}})^T$. The non-zero components of the $5 \times 5$ matrix $H_{{\bf k}}$ are given by $(H_{{\bf k}})_{\tau \tau'}=\exp(i {\bf k}\cdot {\bf v}_{\tau \tau'})$, where ${\bf v}_{\tau \tau'}$ is the vector connecting two NN sites $\tau$ and $\tau'$, which results in the Hamiltonian
\begin{equation}
H_{{\bf k}}=\begin{pmatrix}
0 & e^{-i {\bf k}\cdot {\bf a}_2/2} & 0 & e^{i {\bf k}\cdot {\bf a}_2/2} & 0 \\
e^{i {\bf k}\cdot {\bf a}_2/2} & 0 & e^{i {\bf k}\cdot {\bf a}_1/2} & 0 & e^{i {\bf k}\cdot {\bf a}_3/2} \\
0 & e^{-i {\bf k}\cdot {\bf a}_1/2} & 0 & e^{i {\bf k}\cdot {\bf a}_1/2} & 0 \\
e^{-i {\bf k}\cdot {\bf a}_2/2} & 0 & e^{-i {\bf k}\cdot {\bf a}_1/2} & 0 & e^{-i {\bf k}\cdot {\bf a}_3/2} \\
0 & e^{-i {\bf k}\cdot {\bf a}_3/2} & 0 & e^{i {\bf k}\cdot {\bf a}_3/2} & 0
\end{pmatrix}.
\end{equation}
Diagonalizing $H_{{\bf k}}$ yields the energy bands $E({\bf k})$ depicted in Fig. ~\ref{lattice} (b).

Interestingly, the band structure displays two different regimes. At fillings $f=1/5$ and $f=4/5$, two independent Dirac cones located at ${\bf K}_+=(\frac{2\pi}{3a}, -\frac{2\pi}{3\sqrt{3}a})$ and ${\bf K}_-=(\frac{2\pi}{3a}, \frac{2\pi}{3\sqrt{3}a})$ are present within the first Brillouin zone, which is the same as for the standard honeycomb lattice where Dirac-like dispersion relations effectively describe spin-$1/2$ relativistic fermions.  At half-filling $f=1/2$, a flat band is present at the tip of a single Dirac cone located at  ${\Gamma}_0=(0,0)$. We find that the wavefunctions associated with the flat band have zero amplitude at the green sites ($\tau=1,3,5$) illustrated in Fig. \ref{lattice}(a), which is compatible with the localization property expected from their infinite effective band mass \cite{DOS_T3}. This peculiar configuration, involving a flat band and a single Dirac cone, is also present in the Lieb lattice and leads to an effective Hamiltonian describing spin-$1$ relativistic  fermions ~\cite{shen:2010, apaja:2010,goldmanlieb}. We see from Fig. ~\ref{lattice} (b) that the ECH lattice indeed contains both spin-$1/2$ and spin-$1$ relativistic dispersion relations. These two singular and distinct regimes could be reached in a cold-atom realization by simply tuning the atomic filling factor. In order to further explore the distinction between these two relativistic regimes, we show the density of states $\rho (E)$ in Fig. ~\ref{lattice} (d).  We find that around $E \approx \pm \sqrt{3} t$, $\rho (E)$ behaves linearly which is expected for spin-$1/2$ relativistic fermions. Around $E \approx 0$, the $\rho (E)$ shows a linear behavior and a sharp peak as a consequence of the flat band  \cite{DOS_T3}. As we will demonstrate below, the van Hove singularities located at $E \approx \pm \sqrt{2} t$ constitute natural boundaries separating the spin-1/2 and spin-$1$ relativistic regimes. At this point, let us comment on the interesting fact that the additional (green) lattice sites, i.e., $\tau=1,3,5$, do not destroy the relativistic properties stemming from the background honeycomb lattice. They rather enrich the quantum properties of the lattice in a non-trivial manner by inducing new relativistic regimes at various fillings.

\section{The spin-1/2 and spin-1 relativistic regimes}

To demonstrate the above assertion that the low-energy excitations around fillings $f=1/5, 4/5$ and $f=1/2$, are indeed Dirac-Weyl fermions of different kinds, we obtain the low-energy effective Hamiltonians describing these excitations around the band-touching points, i.e., ${\bf K_{\pm}}$ at $E = \pm \sqrt{3} t$ and ${\Gamma}_0$ at $E = 0$. To do so, we linearize $H({\bf k})$ near ${\bf K}_{\pm}$  or  ${\Gamma}_0$, and subsequently project onto the subspace associated with the two (for $f=1/5$ and $f=4/5$) or three (for $f=1/2$) touching bands. This leads to
\begin{eqnarray}
h_{{\bf p}}^{1/2}&=&\nu_{1/2}(p_x \sigma_1+p_y \sigma_2) \quad {\rm at} \quad f=1/5, 4/5\label{half}\\
h_{{\bf p}}^{1}&=&\nu_{1}(p_x S_1+p_y S_2) \quad  {\rm at} \quad f=1/2\label{integer}
\end{eqnarray} 
where ${\bf p}={\bf k}-{\bf K}_{\pm}({\Gamma}_0)$ and $\nu_{1/2}=\sqrt{3} t/4$, $\nu_{1}=\sqrt{3} t/2$ are the Fermi velocities of the  spin-$1/2$ and spin-$1$ relativistic fermions. We note here that $\nu_{1}= 2 \nu_{1/2}$ is perfectly compatible with their spin-$1$ and spin-$1/2$ nature. While around $f=1/5, 4/5$, the $\sigma_{1,2}$ are the usual Pauli matrices acting on the two-dimensional subspace associated with the two touching bands, the effective Dirac-Weyl Hamiltonian around $f=1/2$ features the $3 \times 3$ matrices $S_{1,2}$, which fulfill the angular momentum commutation relation $[S_i, S_j]=i\epsilon_{ijk}S_k$. Such an effective Hamiltonian has been shown to describe spin-$1$ massless Dirac-Weyl fermions, as recently discussed in Refs.~\cite{shen:2010, apaja:2010,goldmanlieb}.  The spin-$1/2$ and spin-$1$ Dirac-Weyl fermions therefore do indeed {\it coexist} in the ECH lattice at different fillings.

\section{The ``honeycomb-Kagome" decoupling}

\begin{figure}[!hbp]
	\centering
	\includegraphics[width=1\columnwidth]{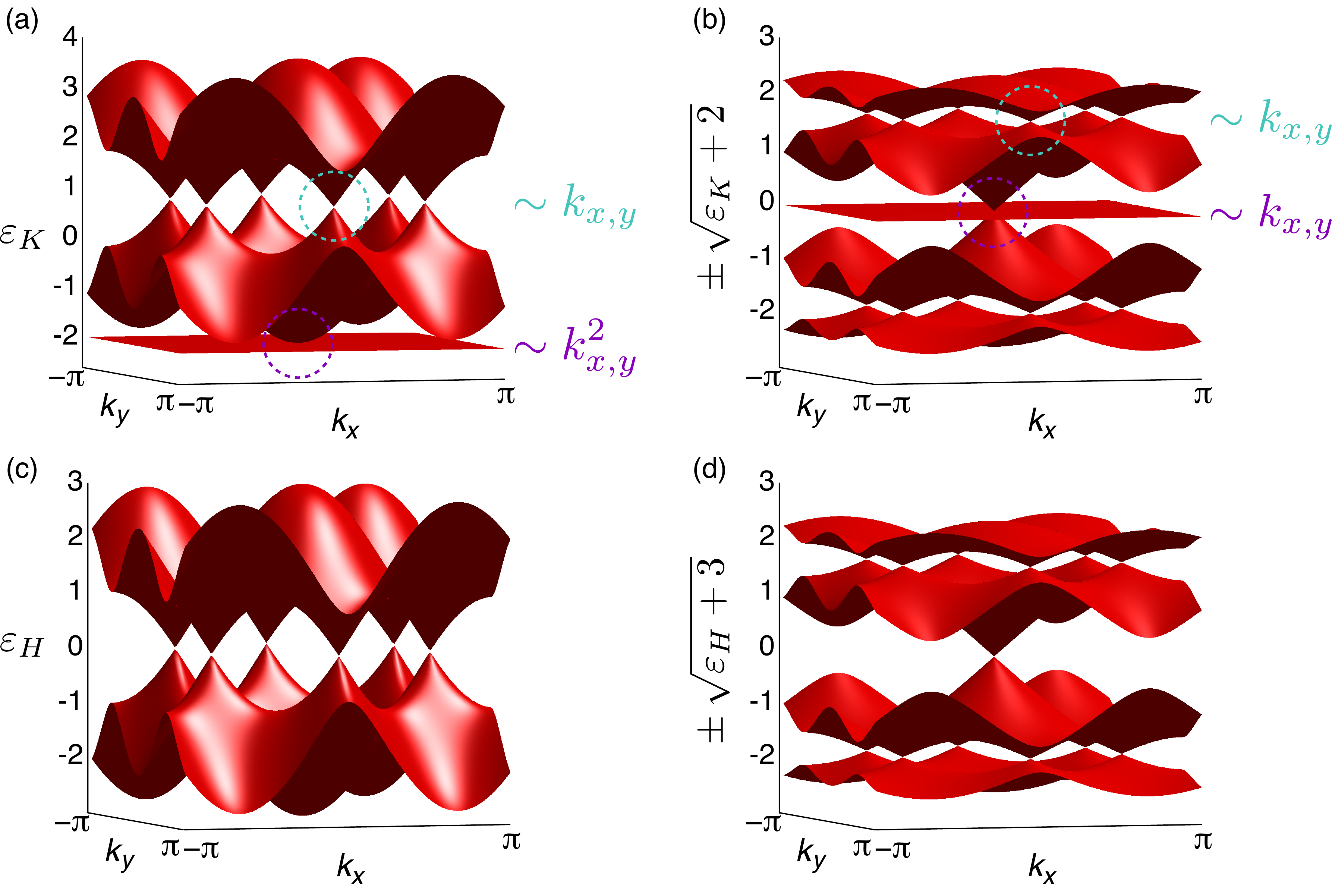}
	\caption{Energy spectrum for the (a) Kagome and (c) Honeycomb lattices. Figures (b),(d) show the mapping \eqref{mymap}-\eqref{mymap2}, leading to the ECH lattice band structure.}
	\label{mapping}
\end{figure}

The ECH lattice has a bipartite structure which allows for an effective decoupling between the red ($\tau=2,4$) and green  ($\tau=1,3,5$) sites illustrated in Fig. ~\ref{lattice} (a). Note that while the red sites  form the background honeycomb lattice, the green sites  constitute a Kagome lattice. It turns out that the band structure depicted in Fig. \ref{lattice} (b) can be deduced from the energy spectra describing these two sublattices \cite{butterfly_T3}. Indeed, as demonstrated in Appendix \ref{myappendix}, one finds that the five energy bands associated to the ECH lattice are directly obtained from the relations 
\begin{align}
&(E/t)(\bs k)=\pm \sqrt{\varepsilon_K (\bs k)+2}, \label{mymap} \\
&(E/t)(\bs k)=\pm \sqrt{\varepsilon_H(\bs k)+3},\label{mymap2}
\end{align}
where $\varepsilon_H (\bs k)$ and $\varepsilon_K (\bs k)$ are the energy bands related to the decoupled honeycomb and Kagome lattices, and where we assume that $E \ne 0$. The band structure $\varepsilon_K (\bs k)$ is illustrated in Fig. \ref{mapping} (a), which shows Dirac points at $\varepsilon_K=1$ and a flat band at $\varepsilon_K=-2$. Note that the dispersion relation is quadratic in the vicinity of the flat band. The band structure of the ECH lattice can then be entirely understood from the spectrum $\varepsilon_K (\bs k)$. From the relation \eqref{mymap}, one obtains a flat band at $E/t=-2+2=0$ and Dirac points at $E/t = \pm \sqrt{1+2}$. Furthermore, the quadratic dispersion of the Kagome lattice around $\varepsilon_K\approx -2$ leads to the conical intersection at $E/t=0$ (see Fig. \ref{mapping} (b)). Therefore both the spin-1 and spin-1/2 Dirac structures stem from the background Kagome lattice, formed by the green sites in Fig. \ref{lattice} (a). However, it is worth emphasizing that the Kagome lattice alone does not display a spin-1 Dirac structure, which highlights the richness of the bipartite ECH lattice with respect to its underlying honeycomb and Kagome lattices. Furthermore, we note that the spectrum associated to the honeycomb lattice $\varepsilon_H (\bs k)$ does not contribute to the band structure $E(\bs k)$ in a significant manner. If we indeed omit the flat band at $\varepsilon_K=-2$, we find that $\pm \sqrt{\varepsilon_K (\bs k)+2}=\pm \sqrt{\varepsilon_H(\bs k)+3}$ (see Fig. \ref{mapping} (c),(d)). 

The ``honeycomb-kagome" decoupling described by Eqs. \eqref{mymap} and \eqref{mymap2} also explains the location of the van Hove singularities in Fig. \ref{lattice} (d). The honeycomb lattice presents van Hove singularities \cite{DOS_Honeycomb} at $\varepsilon_H= \pm 1$, which lead to the four peaks at $E/t = \pm \sqrt{+ 1 +3}$ and $E/t = \pm \sqrt{-1 +3}$ in Fig. \ref{lattice} (d). The linear behavior of the DOS around $E/t \approx \pm \sqrt{3}$ and $E/t \approx 0$ is also easily deduced from the conical intersections stemming from the Kagome lattice, as discussed above.

Finally, we note that the ECH lattice has a bipartite nature, with $N_K=3$ and $N_H=2$ sites per unit cell, where $N_{K,H}$ respectively denotes the number of green (i.e. $\tau=1,3,5$) and red (i.e. $\tau=2,4$) sites. Under such conditions, and since the tunneling only occurs between red and green sites, one can apply the theorem of Ref. \cite{DOS_T3}, which stipulates that a flat band necessarily exists in the energy spectrum and that its weight is given by $N_K/5-N_H/5=1/5$ in the normalized DOS.

\section{ Perturbations} 

We proceed by investigating the effects of various perturbations on the ECH lattice and the two different Dirac-Weyl fermions, highlighting their similar and different natures. We will in particular consider the effects of external magnetic fields, spin-orbit coupling and charge density waves. Such perturbations will not only provide interesting insights into the physical properties of the ECH lattice, but they will also offer the possibility to distinguish between the different relativistic species in an experiment.

\subsection{ Synthetic magnetic fields}  

\begin{figure}[!hbp]
	\centering
	\includegraphics[width=0.7\columnwidth]{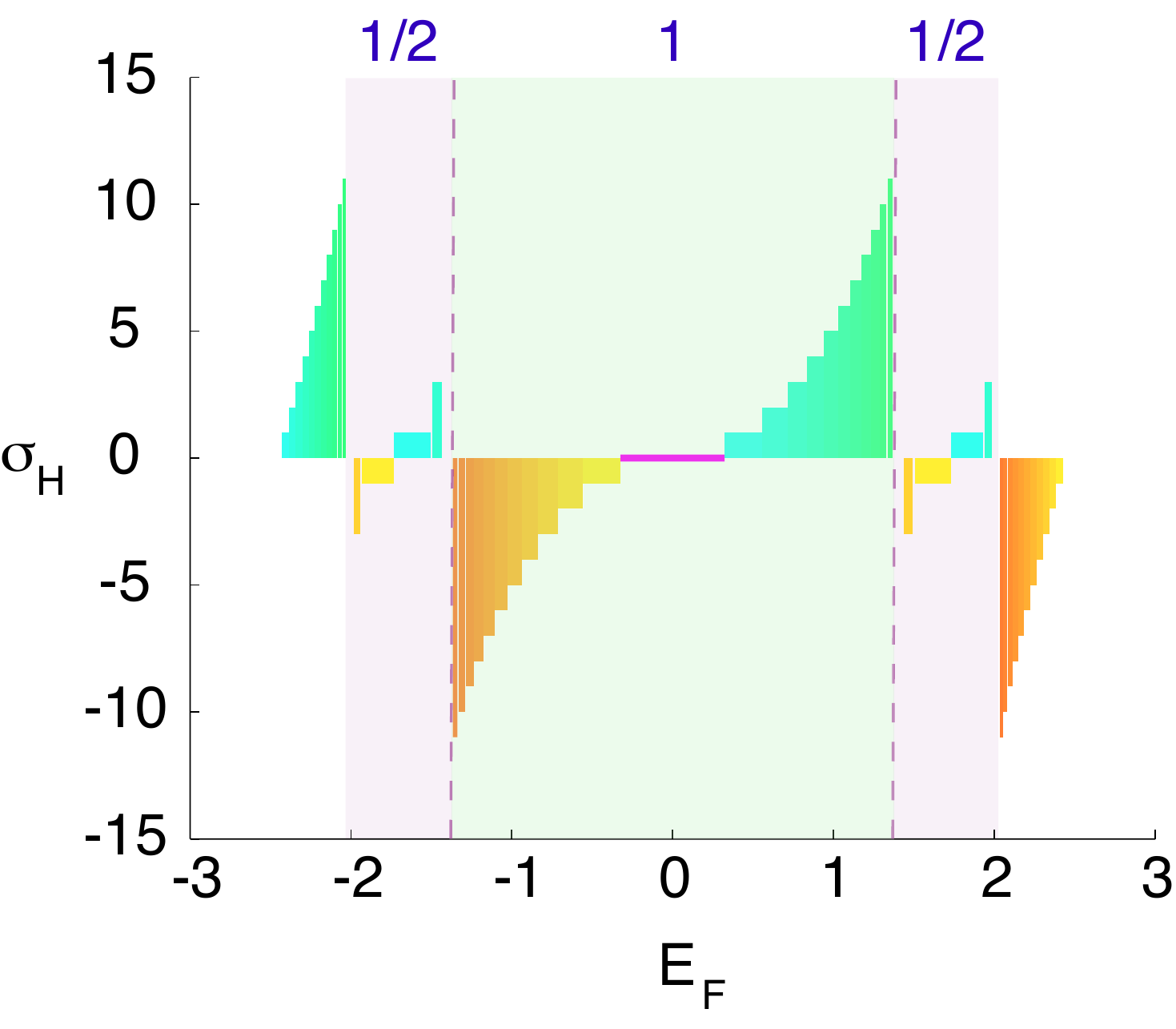}
	\caption{Hall conductivity $\sigma_{\text{H}} (E_{\text{F}})$ as a function of the Fermi energy for $\Phi \approx 0.05$. Vertical dotted lines indicate the location of van Hove singularities at $E/t= \pm \sqrt{2}$ (cf. Fig. ~\ref{lattice} (d)).}
	\label{hall}
\end{figure}

When the ECH lattice is subject to a uniform magnetic flux $\Phi$ per plaquette, the energy bulk gaps form a fractal structure in the $E-\Phi$ plane, which consists of two Hofstadter-Rammal butterflies separated by a flat band~\cite{butterfly}. The butterfly spectra of the honeycomb, Kagome, and $\mathcal{T}_3$ lattices have been reported in Refs~\cite{butterfly_Honeycomb, butterfly_Kagome, butterfly_T3}. When the Fermi energy is exactly located inside such gaps, the Hall conductivity of the system is quantized. A typical sequence of Hall plateaus is shown in Fig.~\ref{hall}, for a reasonably small magnetic flux per plaquette $\Phi \approx 0.05$. The Hall plateaus clearly evolve differently within the different regimes. Around $f=1/5,4/5$, i.e. $E_{\text{F}}= \pm \sqrt{3} t$, the Hall plateaus feature the {\it anomalous double step}  sequence $\sigma_{\text{H}}^{1/2}=\pm 2(N+1/2) e^2/h$, where $N$ is an integer.   In the vicinity of this spin-1/2 regime,  each Dirac fermion contributes to the Hall conductivity according to $\sigma_{\text{Dirac}}=(e^2/h)/2$, i.e., the system exhibits the so-called {\it half-integer} anomalous quantum Hall effect \cite{graphene_hall1,graphene_hall2}. This is also the case for spin-1/2 Dirac fermions in graphene~\cite{graphene_hall}. Around half-filling, i.e., $E_{\text{F}}=0$, one observes the characteristic sequence $\sigma_{\text{H}}^{1}=\pm N_{D} N e^2/h$ describing the quantum Hall plateaus for integer-spin Dirac-Weyl fermions~\cite{weyl_n}, where $N_{\text{D}}$ is the number of Dirac points crossing the flat band. For integer spin Dirac-Weyl fermions, the absence of the {\it half-integer anomaly} leads to a zero Hall conductivity plateau. In Fig. \ref{hall}, we see the characteristic zero Hall conductivity plateau for integer spin Dirac-Weyl fermions and $N_{D}=1$, which is in agreement with the fact that a single Dirac cone is present in this spin-1 regime. 

Interestingly, a sharp change of behavior occurs at $E_{\text{F}}= \pm \sqrt{2} t$ located at the van Hove singularities present in the DOS (see Fig. ~\ref{lattice} (d)), which constitute the boundaries between the $\sigma_{\text{H}}^{1/2}$ and $\sigma_{\text{H}}^{1}$ sequences. We point out that in the standard honeycomb lattice, the van Hove singularities constitute boundaries between relativistic and non-relativistic regimes \cite{graphene_hall2}, which is very different from the result presented here. The Hall conductivity sequence obtained from an ECH lattice subject to a uniform magnetic flux therefore combines the two Hall sequences $\sigma_{\text{H}}^{1/2}$ and $\sigma_{\text{H}}^{1}$ of spin-$1/2$ and spin-$1$ Dirac-Weyl fermions respectively. Obtaining the Hall sequence $\sigma_{\text{H}} (E_{\text{F}})$, such as presented in Fig. \ref{hall}, would provide a clear signature for the coexistence of spin-$1/2$ and spin-$1$ Dirac-Weyl fermions in the ECH lattice. 

In a cold-atom framework, such a study would require the presence of a uniform synthetic magnetic field within the optical ECH lattice. This difficult, but realistic, task would require to engineer Peierls phases $\exp (i \phi_{j})$ that accompany the hopping of the atoms along the links $j$, in such a way that the total product of the phases along a plaquette yields $\prod_{\square} \phi_j=\Phi$, where $\Phi$ is the magnetic flux per plaquette. Such phases could be induced by means of Raman-assisted tunneling (see Ref.~\cite{gauge_field} for a review of synthetic gauge fields for cold atoms), as recently demonstrated experimentally in Ref.~\cite{lattice_gauge}. Signatures related to the Hall sequences could then be obtained from density measurements, as discussed in Ref.~\cite{oktel,non_abelian_honeycomb}.

\subsection{Spin-orbit coupling}  

An intrinsic spin-orbit (SO) coupling term, 
\begin{equation}
H_{SO}=i\lambda_{SO}\sum_{\langle\langle ij \rangle\rangle\alpha\beta}(\boldsymbol{d}^1_{ij}\times \boldsymbol{d}^2_{ij})\cdot \boldsymbol{\sigma}_{\alpha\beta}c^{\dagger}_{i\alpha}c_{j\beta},\label{so}\end{equation}
has been introduced in Ref. \cite{Kane_Mele} to predict the quantum spin Hall effect in a model described by a standard honeycomb lattice. Here $\lambda_{SO}$ is the SO coupling strength, ${\bf d}^1_{ij}$ and ${\bf d}^2_{ij}$ are two vectors connecting the NNN sites $i$ and $j$, and ${\bf \sigma}$ is the vector of Pauli matrices acting on the spin. This term opens a bulk gap in the energy spectrum $\epsilon_H (\bs k)$ that is associated to a non-trivial $Z_2$ index and hosts topologically protected helical edge states (i.e., counter-propagating edge states with opposite spin)~\cite{Kane_Mele}. This NNN SO term has been generalized to other lattices exhibiting spin-$1/2$ relativistic fermions \cite{kagome_TI,decorated_honeycomb_TI}, and systematically leads to non-trivial $Z_2$ phases. In lattices featuring effective spin-$1$ fermions, such as the Lieb or $\mathcal{T}_3$ lattices, the situation is more subtle. The NNN SO term leads to a trivial phase for the $\mathcal{T}_3$ lattice but a non-trivial phase for the Lieb lattice ~\cite{goldmanlieb, lieb_TI,weyl_1}. Therefore, the effect of the SO term on the ECH lattice, in which spin-$1$ and spin-$1/2$ excitations coexist, is {\it a priori} a non-trivial problem. Let us comment on the fact that the NNN hopping defined on the ECH lattice leads to a Kagome structure. However, the path-dependent phases associated to the hopping in Eq. \eqref{so} generates a radically different spectrum, featuring non-trivial bulk gaps (cf. below). Finally, we mention that the SO term \eqref{so} could, in principle, be engineered in optical lattices (cf. Refs. \cite{na1,na2,na3,na4}).

\begin{figure}[!hbp]
	\centering
	\includegraphics[width=0.7\columnwidth]{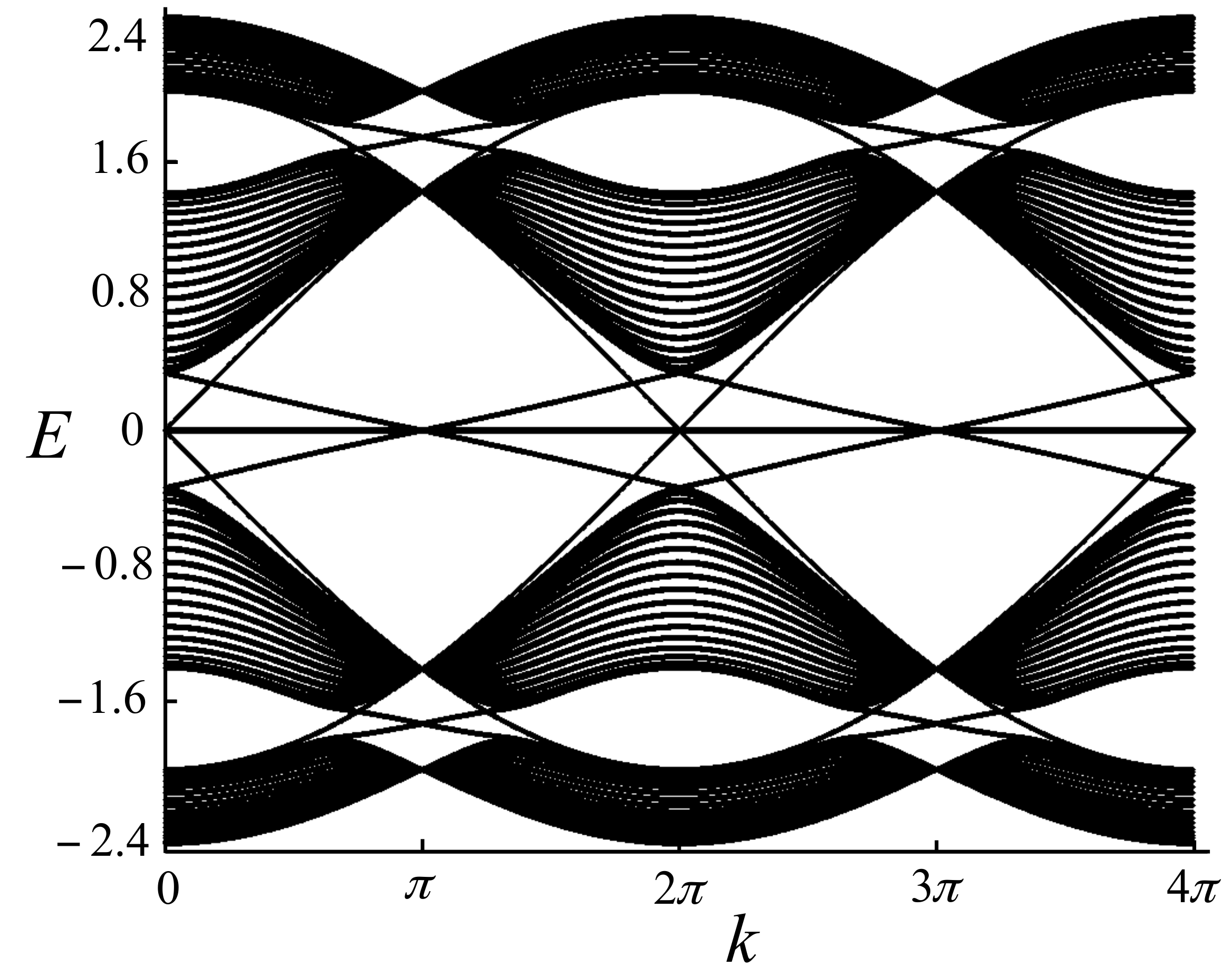}
	\caption{The energy spectrum $E=E(k)$, for an ECH lattice subject to the Haldane-type SO coupling \eqref{so} with periodic boundary conditions along one spatial direction, shows the presence of helical edge states in the vicinity of both the spin-$1/2$ and the spin-$1$ regime.}
\label{edge}
\end{figure}

First, we show how the SO term affects the low-energy theory describing the two kinds of Dirac-Weyl fermions. In this limit the effective Hamiltonians are
\begin{eqnarray}
h_{{\bf k}}^{1/2}&=&\nu_{1/2}(k_x \sigma_1+k_y \sigma_2)-\frac{\sqrt{3}}{2}\alpha\lambda_{SO}  \sigma_3 \label{soone}\\
h_{{\bf k}}^{1}&=&\nu_{1}(k_x S_1+k_y S_2)-2\sqrt{3}\alpha\lambda_{SO}  S_3
\end{eqnarray}
where $\alpha=\pm$ is the spin index. We note that Eq. \eqref{soone} holds for all the spin-$1/2$ Dirac species, namely for all ${\bf K}_{\pm}$ at $f=1/5,4/5$. Therefore, the SO term generates the same mass term for all the spin-$1/2$ relativistic excitations, and thus opens bulk gaps at the four independent spin-$1/2$ Dirac points. Two bulk gaps also appear in the vicinity of the spin-$1$ Dirac point (i.e., $f=1/2$) with the flat band being preserved by the SO term.  Our result indicates that the bulk gaps associated with the spin-$1$ Dirac-Weyl fermion are much bigger than the gap associated with the spin $1/2$ Dirac-Weyl fermions (cf. also Fig.  ~\ref{edge}).

\begin{table}
\begin{tabular}{l|rlrlrlrlrlcl}
		\hline 
			\hline 				
		 $\mathcal{P}_3$ & ${\Gamma}_0$  & ${\Gamma}_1$ & ${\Gamma}_2$ & ${\Gamma}_3$ & $\prod_i$ & $\quad \nu$\\
		
		\hline
			\hline 				
	Band 4 & $+1$ & $-1$ & $+1$ & $-1$ & 1 &$\nu_4=1$\\
				\hline
 Band 3 & $-1$ & $+1$& $-1$ &$+1$ & 1 &$\nu_3=1$\\
	  		\hline
  Band 2 & $+1$ & $-1$& $+1$ &$-1$ & 1 &$\nu_2=1$\\
	  		\hline
 Band 1 & $+1$ & $+1$& $-1$ &$+1$ & -1 & $\nu_1=1$\\
		\hline		
			\hline 
		\end{tabular}
		\caption{ Parity-eigenvalue pattern at the four $\mathcal{T}$-invariant momenta ${\Gamma}_i$ for the four different occupied bands. All the bulk gaps are associated with a non-trivial $Z_2$ index $\nu=1$.}
		\label{parity_pattern}
\end{table}

We now compute the four $Z_2$ indices $\nu_{N}$, with $N=1,2,3,4$ associated with the four bulk gaps opened by $H_{SO}$. Since the ECH lattice possesses inversion symmetry, the $Z_2$ topological invariant $\nu_{N}$ associated with the $N$th bulk gap can be easily evaluated through the formula ~\cite{inversion1, inversion2}
\begin{equation} 
\prod_{i=0}^3 \prod_{m=1}^{N} \xi_{2m}({\Gamma}_i)=(-1)^{\nu_N}. \label{inversion}
\end{equation} 
In this expression,  $\xi_{2m}({\Gamma}_i) =\pm 1$ is the parity eigenvalue associated with the $2m$-th occupied energy band, which is evaluated at one of the four $\mathcal{T}$-invariant momenta $\bs k={\Gamma}_i$.  The latter can be expressed as ${\Gamma}_i={\bf \hat{q}}_1 n_i/2+{\bf \hat{q}}_2 m_i/2$ with $n_i=\{0,1\}$ and $m_i=\{0,1\}$, where ${\bf \hat{q}}_1=2\pi/3(1,\sqrt{3})$ and ${\bf \hat{q}}_2=2\pi/3(1,-\sqrt{3})$. Choosing the site $\tau=3$ inside the unit cell as the center of inversion, the parity operator acts as 
\begin{align}
&\mathcal{P}_3[\psi_1({\bf r}), \psi_2({\bf r}),\psi_3({\bf r}),\psi_4({\bf r}),\psi_5({\bf r})] \nonumber \\
&=[\psi_1({\bf -r}+{\bf a}_1-{\bf a}_2), \psi_4(-{\bf r}+{\bf a}_1),\psi_3(-{\bf r}), \nonumber \\
& \qquad \qquad \qquad \qquad \qquad  \psi_2({\bf -r}-{\bf a}_1),\psi_5({\bf -r}+{\bf a}_3-{\bf a}_1)],
\end{align}
where $\psi_{\tau} ({\bf r})$ is the single-particle wavefunction defined at site $\tau$. The eigenstates of $H_{0+SO}(\bs k=\Gamma_i) $, as well as the parity eigenvalues of the occupied bands, are determined numerically, yielding the results presented in Table~ \ref{parity_pattern}. We find that the expression \eqref{inversion} gives $(-1)^{\nu}=-1$ for each bulk gap, indicating that the $Z_2$ phases generated by the SO term are all non-trivial. Therefore, the spin-1 regime of the ECH lattice behaves similarly to the Lieb lattice ~\cite{goldmanlieb, lieb_TI,weyl_1}. 

To further confirm these results, we diagonalize the ECH lattice in the presence of the SO term and consider periodic boundary conditions along one spatial direction. In this cylindrical geometry, the energy spectrum features helical edge states within the four bulk gaps predicted by the non-trivial $Z_2$ index  \cite{Kane_Mele,kagome_TI,decorated_honeycomb_TI} (see Fig. \ref{edge}). We note that the dispersion relations $E(k)$ associated with the edge states at half-filling are similar to those obtained from the Lieb lattice ~\cite{goldmanlieb, lieb_TI}, further illustrating the similarity between the spin-$1$ regimes of the ECH  and Lieb lattices. In the vicinity of the spin-$1/2$ regime, i.e., $f=1/5, 4/5$, the dispersion relation of the edge states is similar to those obtained in the Kagome lattice \cite{kagome_TI}.  These results show that the SO term acts in a non-trivial way, both for the spin-$1$ and spin-$1/2$ regimes, indicating that the ECH lattice presents striking similarities with both the Kagome and the Lieb lattices. However, the presence of  a SO term does not allow us to distinguish between the spin-$1$ and spin-$1/2$ regimes, as they would both give rise to the same quantum spin Hall effect. Finally, we mention that edge states structures, such as depicted in Fig. \ref{edge}, could be probed in atomic systems through Bragg spectroscopy \cite{bragg2,bragg,bragg3}.

\subsection{Charge density waves (CDW)} 

In the honeycomb lattice, a staggered potential is known to open a trivial bulk gap at half-filling \cite{Kane_Mele}. Such a perturbation acts as local chemical potentials $\mu_A=-\mu_B$, which take opposite values at the sites $A$ and $B$ constituting the unit cell of the honeycomb.  Such a perturbation has been generalized for lattices featuring $\mathcal{N}>2$ sites per unit cell, such as the Kagome~\cite{kagome_TI} and the decorated honeycomb lattice ~\cite{decorated_honeycomb_TI}. This charge-density-wave term is expressed as $H^{CDW}_{\boldsymbol{k}}=\text{diag} (\mu_1 , \dots \mu_{\mathcal{N}})$, which reduces to the honeycomb staggered potential for $\mathcal{N}=2$ and $\mu_1=-\mu_2$. For lattices with $\mathcal{N}>2$ exhibiting spin-1/2 Dirac fermions, it was shown that this CDW takes the form of an axial gauge field in the effective low-energy Hamiltonian ~\cite{kagome_TI, decorated_honeycomb_TI}, namely a gauge potential $\boldsymbol{A}$ which has opposite sign at two independent Dirac points.

Here we are interested in the fate of the spin-$1/2$ and spin-$1$ Dirac points when a charge density wave term $H^{CDW}_{{\bf k}}=t \, \text{diag} (\mu_1 , \dots ,\mu_5)$ is added in the ECH lattice, where the local chemical potentials $\mu_{\tau}$ can be individually tuned.  In the vicinity of the four spin-1/2 Dirac points $K_{\pm}$, we find that the low-energy terms corresponding to the CDW take the form
\begin{eqnarray}
h({\bf p})_{CDW}^{1/2}&=&-(A_x^l \sigma_1+A_y^l \sigma_2-A_z^l \sigma_3)+A_0 I, \quad f=\frac{1}{5} \\ 
h({\bf p})_{CDW}^{1/2}&=&(A_x^l \sigma_1+A_y^l \sigma_2+A_z^l \sigma_3)+A_0 I, \quad f=\frac{4}{5}\label{cdw}
\end{eqnarray}
where $A_x^l=(\mu_1-\mu_5)l/4\sqrt{3}$,  $A_y^l =(\mu_1-2\mu_3+\mu_5)l/12$,  $A_z^l =(\mu_2-\mu_4)l/4$, $A_0 =(2\mu_1+3\mu_2+2\mu_3+3\mu_4+2\mu_5)/12$, and $l=\pm$ refers to the two Dirac points ${\bf K_{\pm}}$. 
Therefore, when $\mu_2=\mu_4$, i.e., $A_z^l =0$, and similar to the results reported for the Kagome lattice, we find that the CDW acts as an axial gauge field. In other words, the low-energy Hamiltonians 
\begin{equation}
h({\bf p})^{1/2}=\sum_{\nu} v_{1/2} (k_{\nu}- \mathcal{A}_{\nu}^l) \sigma_{\nu},
\end{equation}
where $ \mathcal{A}_{\nu}^l=A_{\nu}^l/v_{1/2}$ and $\mathcal{A}_{\nu}^{+}=-\mathcal{A}_{\nu}^{-}$, can be expressed in terms of a gauge potential ${\bf A}$ which has opposite sign at two independent Dirac points.
In this case the effect of the gauge potential $\mathcal{A}_{\nu}^{\pm}$ on the spin-$1/2$ Dirac-Weyl fermions is to move the positions of their Dirac points inside the Brillouin zone. At a given filling, the displacement of the two Dirac points are opposite, since the gauge field is axial. Furthermore, these displacements are in opposite directions at fillings $f=1/5$ and $f=4/5$ (cf. Eq. \eqref{cdw}). In other words, if two Dirac cones come closer at $f=1/5$ as the CDW is increased, the two cones at $f=4/5$ will separate. When the CDW is sufficiently strong, the two approaching Dirac points at $f=1/5$ will annihilate each other~\cite{non_abelian_honeycomb, merging2,moving_dirac_points}, while the two Dirac points at $f=4/5$ will survive. This process allows the destruction of a pair of spin-$1/2$ species at a given filling while preserving the others.

In contrast, by expanding the Hamiltonian around the spin-$1$ Dirac point at $\Gamma_0$, we find that the low-energy form of the CDW perturbation is more involved than for the spin-$1/2$ regime, and that it cannot be simply interpreted as a gauge field. Indeed, the low-energy limit of the CDW term cannot be written as a superposition of the three angular-momentum matrices $S_{x, y, z}$ which do not form a complete basis for $3 \times 3$ matrices. This interesting result indicates that the spin-$1/2$ and spin-$1$ regimes of the ECH lattice should react differently to the CDW, and thus providing a mechanism for  distinguishing them in an experiment. 

\begin{figure}
	\centering
	\includegraphics[width=1.\columnwidth]{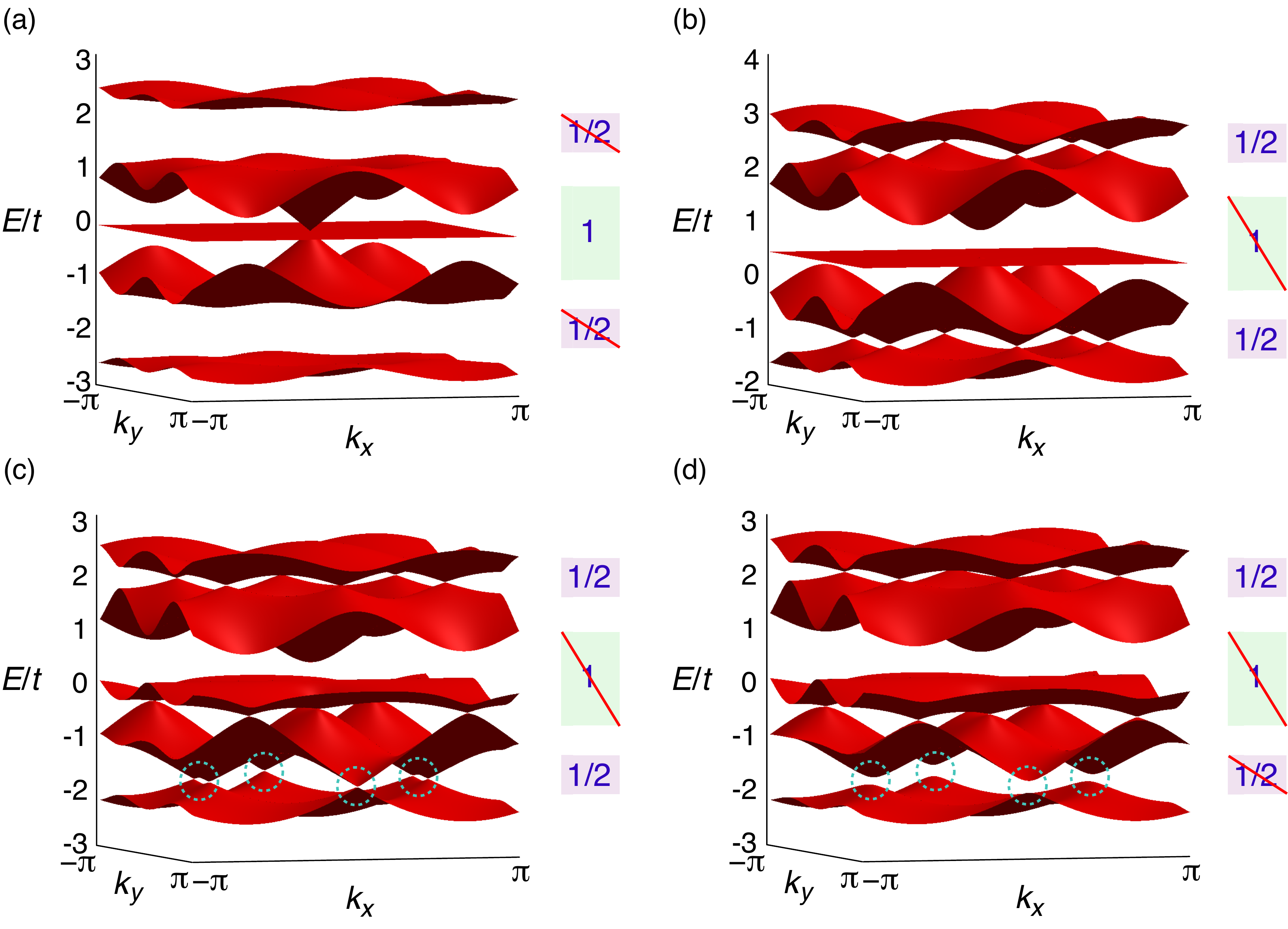}
	\caption{Energy bands $E=E(k_x,k_y)$ for different configurations of the CDW: (a) $\mu_2=-\mu_4=1$ and $\mu_{1,3,5}=0$: the spin-1/2 regimes are gapped while the spin-1 regime is preserved. (b) $\mu_2=\mu_4=1$ and $\mu_{1,3,5}=0.5$: the spin-1 regime is gapped while the spin-1/2 regimes and the flat band are preserved. (c) $\mu_2=\mu_4=0.5$, $\mu_1=0.25$, $\mu_3=-0.6$ and $\mu_5=0.1$: the spin-1 regime is gapped, the flat band is perturbed, and the robust Dirac points at $f=1/5$ move closer (cf. dotted circles). (d) The same CDW configuration as in (c) but multiplied by a factor $1.4$ (i.e. $\mu_2=\mu_4=0.7$, $\mu_1=0.35$, $\mu_3=-0.84$ and $\mu_5=0.14$ ): the Dirac points at $f=1/5$ annihilate each other (cf. dotted circles), and only the spin-1/2 Dirac points at $f=4/5$ survive.}
	\label{cdwfig}
\end{figure}

The possibility to destroy and preserve the spin-1/2 and spin-1 fermions \emph{individually}, using the CDW perturbation, is appealing. Here we report a selection of relevant configurations that achieve this goal.
\begin{itemize}
\item $\mu_2=-\mu_4$ and $\mu_1=\mu_3=\mu_5$. In this case, the CDW acts as a staggered potential for the background honeycomb lattice. It destroys all the spin-1/2 Dirac points at $\boldsymbol{K}_{\pm}$ and $f=1/5,4/5$ by opening trivial bulk gaps \cite{Kane_Mele}. By setting $\mu_1=\mu_3=\mu_5$, the CDW do not perturb the localized states defined at the green sites ($\tau=1,3,5$) of Fig. 1 (a). The flat band is therefore preserved. In addition, we find that when $\mu_2=-\mu_4$, the spin-1 Dirac point at $\Gamma_0$ only survives for $\mu_{1,3,5}=0$. This situation is illustrated in Fig. \ref{cdwfig} (a).
\item $\mu_2=\mu_4$ and $\mu_1=\mu_3=\mu_5$. In this case, $A_{x}^l=A_{y}^l=A_{z}^l=0$ and $A_{0}^l \ne 0$, thus the spin-1/2 Dirac points are simply shifted in energy. Since $\mu_1=\mu_3=\mu_5$, the flat band is preserved but the spin-1 Dirac point at $\Gamma_0$ is generally destroyed. This situation is illustrated in Fig. \ref{cdwfig} (b).
\item $\mu_2=\mu_4$ and arbitrary $\mu_{1,3,5}$. In this case, the CDW acts as a non-trivial axial gauge field and the spin-1/2 Dirac points move inside the Brillouin zone in opposite directions.
Therefore, for small CDW, the spin-1/2 fermions are all preserved (cf. Fig. \ref{cdwfig} (c)). For larger CDW, two fermions generally annihilate each other at $f=1/5$ \emph{or} $f=4/5$ (the displacements being in opposite direction at these fillings), in which case only one spin-1/2 regime survives (cf. Fig. \ref{cdwfig} (d)). In addition, for arbitrary $\mu_{1,3,5}$ the flat band and the spin-1 Dirac fermion are generally destroyed. 
\end{itemize}

Therefore, by selecting the appropriate configuration of the CDW perturbation, one is able to engineer a system whose band structure displays zero, one or two spin-1/2 Dirac regimes, a flat band or not, a spin-1 Dirac regime or not. 

\section{ Conclusions} 

We have investigated the ECH lattice which features both spin-$1/2$ and spin-$1$ Dirac-Weyl fermions at different filling fractions. By using an intriguing mapping, we have shown that the underlying Kagome and honeycomb structures of the ECH lattice play a crucial role in determining the band structure of the ECH lattice. We have also explored several types of perturbations on the lattice which offer a powerful control over this rich system. It is certainly tempting to extend this scenario to include collisional interactions between the spins. This will not only allow for exotic new phases ~\cite{honeycomb_interaction, Lieb_interaction} and applications of models such as the Kitaev anyonic model~\cite{anyon}, but will also hopefully shed light on open questions at the forefront of condensed-matter physics.

\acknowledgements 

ZL acknowledges the  support from SUPA (Scottish Universities Physics Alliance). NG thanks the FRS-FNRS for financial support. P\"O acknowledges support from the Carnegie Trust for the Universities of Scotland.

\appendix

\section{The single-particle Schr\" odinger equation and the ``honeycomb-kagome" mapping}
\label{myappendix}

The single-particle Schr\"odinger equation describing non-interacting particles on the ECH lattice can be directly derived from the second-quantized Hamiltonian \eqref{myham}. Denoting the wave function at lattice site $\tau=1, \dots , 5$ by $\psi_{\tau} (\bs{x})$, with spatial coordinate $\bs{x}$, one finds the set of coupled equations
\begin{widetext}
\begin{align}
&(E/t) \psi_1 (\bs x)= \psi_4 (\bs x + \bs{a}_2/2)+\psi_2 (\bs x - \bs{a}_2/2), \label{apeq} \\ 
&(E/t) \psi_2 (\bs x - \bs{a}_2/2)= \psi_1 (\bs x)+\psi_3 (\bs x - \bs{a}_2/2 +\bs{a}_1/2)+\psi_5 (\bs x - \bs{a}_2/2 +\bs{a}_3/2),  \nonumber \\
&(E/t) \psi_3 (\bs x - \bs{a}_2/2 +\bs{a}_1/2)= \psi_4 (\bs x - \bs{a}_2/2+\bs{a}_1)+\psi_2 (\bs x - \bs{a}_2/2),  \nonumber \\
&(E/t) \psi_4 (\bs x - \bs{a}_2/2+\bs{a}_1)= \psi_1 (\bs x  - \bs{a}_2 +\bs{a}_1)+\psi_3 (\bs x - \bs{a}_2/2 +\bs{a}_1/2)+\psi_5 (\bs x - \bs{a}_2/2 -\bs{a}_3/2 +\bs{a}_1),  \nonumber \\
&(E/t) \psi_5 (\bs x - \bs{a}_2/2 -\bs{a}_3/2 +\bs{a}_1)= \psi_2 (\bs x - \bs{a}_2/2 +\bs{a}_1 -\bs{a}_3)+\psi_4 (\bs x - \bs{a}_2/2 +\bs{a}_1),  \nonumber
\end{align}
\end{widetext}
where ${\bf a}_1=(1,0)$, ${\bf a}_2=(-1/2,\sqrt{3}/2)$, ${\bf a}_3=(-1/2,-\sqrt{3}/2)$ (see main text). For $E \ne 0$, one can decouple  \eqref{apeq} into two subsets of equations describing the red (i.e., $\tau=2,4$) and green sites  (i.e., $\tau=1,3,5$) independently. We find
\begin{widetext}
\begin{align}
&\bigl( (E/t)^2 -2 \bigr )\psi_1 (\bs x)= \psi_3 (\bs x + \bs{a}_2/2 - \bs{a}_1/2)+\psi_3 (\bs x - \bs{a}_2/2 + \bs{a}_1/2)+\psi_5 (\bs x + \bs{a}_2/2 - \bs{a}_3/2) +\psi_5 (\bs x - \bs{a}_2/2 + \bs{a}_3/2),  \\ 
&\bigl( (E/t)^2 -2 \bigr )\psi_3 (\bs x - \bs{a}_2/2 + \bs{a}_1/2)= \psi_1 (\bs x)+\psi_1 (\bs x+ \bs{a}_1 - \bs{a}_2)+ \psi_5 (\bs x - \bs{a}_2/2 + \bs{a}_3/2)+ \psi_5 (\bs x - \bs{a}_2/2 - \bs{a}_3/2 + \bs{a}_1),   \nonumber \\
&\bigl( (E/t)^2 -2 \bigr )\psi_5 (\bs x - \bs{a}_2/2 - \bs{a}_3/2 + \bs{a}_1)=\psi_1 (\bs x+ \bs{a}_1 - \bs{a}_2)+\psi_1 (\bs x+ \bs{a}_1 - \bs{a}_3) +\psi_3 (\bs x - \bs{a}_2/2 + \bs{a}_1/2)+\psi_3 (\bs x - \bs{a}_2/2 + 3 \bs{a}_1/2 - \bs{a}_3),   \nonumber \\
\end{align}
and 
\begin{align}
&\bigl( (E/t)^2 -3 \bigr )\psi_2 (\bs x - \bs{a}_2/2)= \psi_4 (\bs x + \bs{a}_2/2) + \psi_4 (\bs x - \bs{a}_2/2 + \bs{a}_1)+\psi_4 (\bs x - \bs{a}_2/2 + \bs{a}_3),  \\ 
&\bigl( (E/t)^2 -3 \bigr )\psi_4 (\bs x - \bs{a}_2/2 + \bs{a}_1)= \psi_2 (\bs x - \bs{a}_2/2) + \psi_2 (\bs x - 3 \bs{a}_2/2 + \bs{a}_1)+\psi_2 (\bs x - \bs{a}_2/2 + \bs{a}_1 - \bs{a}_3).   \nonumber 
\end{align}
\end{widetext}
Writing $\psi_{\tau} (\bs x)= \exp (i \bs k \cdot \bs x) \phi_{\tau}$, one finds the two separate eigensystems
\begin{widetext}
\begin{equation}
\varepsilon_K  (\bs k) \begin{pmatrix} \phi_1 \\ \phi_3 \\ \phi_5 \end{pmatrix}=
2 \begin{pmatrix}
0 & \cos \bs k \cdot (\bs a_2 - \bs a_1)/2 & \cos \bs k \cdot (\bs a_2 - \bs a_3)/2 \\ 
\cos \bs k \cdot (\bs a_2 - \bs a_1)/2 & 0 & \cos \bs k \cdot (\bs a_3 - \bs a_1)/2 \\
\cos \bs k \cdot (\bs a_2 - \bs a_3)/2 & \cos \bs k \cdot (\bs a_3 - \bs a_1)/2 & 0
\end{pmatrix}\begin{pmatrix} \phi_1 \\ \phi_3 \\ \phi_5 \end{pmatrix}= H_{K} (\bs k) \begin{pmatrix} \phi_1 \\ \phi_3 \\ \phi_5 \end{pmatrix},
\end{equation}
and 
\begin{equation}
\varepsilon_H  (\bs k) \begin{pmatrix} \phi_2 \\ \phi_4 \end{pmatrix}=
\begin{pmatrix}
0 & \sum_{\nu} e^{i \bs k \cdot \bs a_{\nu}}\\ 
\sum_{\nu} e^{-i \bs k \cdot \bs a_{\nu}} & 0
\end{pmatrix}\begin{pmatrix} \phi_2 \\ \phi_4 \end{pmatrix}= H_{H} (\bs k) \begin{pmatrix} \phi_2 \\ \phi_4 \end{pmatrix},
\end{equation}
\end{widetext}
where we introduced the dimensionless quantities 
\begin{equation}
\varepsilon_K  (\bs k)=(E (\bs k)/t)^2-2, \qquad \varepsilon_H  (\bs k)=(E (\bs k)/t)^2-3. \label{echmap}
\end{equation}
The two decoupled systems, described by the Hamiltonians $H_{H} (\bs k)$ and $H_{K} (\bs k)$, correspond to the honeycomb and Kagome lattices formed by the red (i.e., $\tau=2,4$) and green (i.e., $\tau=1,3,5$) sites, respectively. The energy bands of the two subsystems are given by 
\begin{align}
&\varepsilon_K (\bs k)=(1\pm\sqrt{4A_{\bf k}-3}), \qquad \varepsilon_K (\bs k)=-2, \label{kagomemap}\\
&\varepsilon_H= \pm \vert  \sum_{\nu=1}^3 e^{i \bs k \cdot \bs a_{\nu}} \vert \label{honeymap},
\end{align}
where $A_{\bf k}=\cos^2[{{\bf k}\cdot ({\bf a}_2-{\bf a}_3)/2}]+\cos^2[{{\bf k}\cdot ({\bf a}_3-{\bf a}_1)/2}]+\cos^2[{{\bf k}\cdot ({\bf a}_1-{\bf a}_2)/2}]$. We note that $\sqrt{4A_{\bf k}-3}=\vert  \sum_{\nu=1}^3 e^{i \bs k \cdot \bs a_{\nu}} \vert$,  thus we find that $\pm \sqrt{\varepsilon_K (\bs k)+2}=\pm \sqrt{\varepsilon_H(\bs k)+3}$ (for $E \ne 0$). The latter result shows that the  band structure describing the ECH lattice, $E(\bs k)\ne 0$, can be equally obtained from the Kagome spectrum $\varepsilon_K (\bs k)$ through the relations \eqref{echmap}-\eqref{kagomemap}, or from the honeycomb spectrum $\varepsilon_H (\bs k)$ through the relations \eqref{echmap}-\eqref{honeymap}. Besides, we note that the flat band at $E=0$ can also be deduced from the Kagome subsystem through \eqref{echmap}-\eqref{kagomemap}, although we  stress that this mapping is only strictly valid for $E \ne 0$. The band structures $\varepsilon_{K,H} (\bs k)$ are illustrated in Figs. \ref{mapping} (a),(c). The five energy bands associated to the ECH lattice, obtained through the relations $E/t=\pm \sqrt{\varepsilon_K+2}$ and $E/t=\pm \sqrt{\varepsilon_H+3}$, are depicted in Figs. \ref{mapping} (b),(d).


\begin{references}


\bibitem{graphene_rev}
A. H. Castro Neto, F. Guinea, N. M. R. Peres, K. S. Novoselov, and A. K. Geim,  Rev. Mod. Phys. {\bf 81}, 109 (2009).

\bibitem{top_insulators_rev} X.-L. Qi and S.-C. Zhang, Rev. Mod. Phys. {\bf 83}, 1057 (2011).

\bibitem{moving_dirac_points} L. Tarruell, D. Greif, T. Uehlinger, G. Jotzu, and T. Esslinger, Nature \textbf{483}, 302 ( 2012). 

\bibitem{Lim2008} L.-K. Lim, C. Morais Smith and A. Hemmerich, Phys. Rev. Lett.  {\bf 100} 130402 (2008).

\bibitem{Goldman2009}  N. Goldman, A. Kubasiak, A. Bermudez, P. Gaspard, M. Lewenstein, and M. A. Martin-Delgado, Phys. Rev. Lett. {\bf 103} 035301 (2009).
\bibitem{Bermudez2010prl} A. Bermudez, L. Mazza, M. Rizzi, N. Goldman, M. Lewenstein, and M. A. Martin-Delgado, Phys. Rev. Lett. {\bf 105} 190404 (2010).

\bibitem{Lee2009} K. L. Lee, B. Gr�maud, R. Han, B.-G. Englert and C. Miniatura, Phys. Rev. A.  {\bf 80}, 043411 (2009).

\bibitem{na2} L. Mazza, A. Bermudez, N. Goldman, M. Rizzi, M.-A. Martin-Delgado and M. Lewenstein, New J. Phys. {\bf 14}  015007 (2012).


\bibitem{weyl_1}
D. Bercioux, D. F. Urban, H. Grabert, and W. Häusler,  Phys. Rev. A {\bf 80},  063603 (2009); D. Bercioux, N. Goldman, and D. F. Urban, Phys. Rev. A {\bf 83}, 023609 (2011).

\bibitem{goldmanlieb} N. Goldman, D. F. Urban and D. Bercioux, Phys. Rev. A {\bf 83}, 063601 (2011)
\bibitem{apaja:2010} V. Apaja, M. Hyrk\"as and M. Manninen, Phys. Rev. A  \textbf{82},   041402(R) (2010). 
\bibitem{shen:2010} R. Shen, L. B. Shao, Baigeng Wang, and D. Y. Xing, Phys. Rev. B \textbf{81},  041410(R) (2010).

\bibitem{weyl_n} Z. Lan, N. Goldman, A. Bermudez, W. Lu, and P.  \"Ohberg, Phys. Rev. B {\bf 84}, 165115 (2011).



\bibitem{lattice_model} H Watanabe, Y Hatsugai, and H Aoki, \textrm{arXiv:1009.1959} (2010); B. D\'ora, J. Kailasvuori, and R. Moessner, Phys. Rev. B {\bf 84}, 195422 (2011).


\bibitem{DOS_T3} B. Sutherland,  Phys. Rev. B \textbf{34},  5208 (1986).



\bibitem{kagome_TI} H.-M.Guo and M. Franz, Phys. Rev. B \textbf{80},  113102 (2009).

\bibitem{Kane_Mele} C. L. Kane and E. J. Mele,  Phys. Rev. Lett. \textbf{95},  146802 (2005).


\bibitem{whyte2005} G. Whyte and J. Courtial, New J. Phys. {\bf 7}, 117 (2005).

\bibitem{artificial_lattice} A. Singha, M. Gibertini, B. Karmakar, S. Yuan, M. Polini, G. Vignale, M. I. Katsnelson, A. Pinczuk, L. N. Pfeiffer, K. W. West, and V. Pellegrini,  Science \textbf{332,} 1176 (2011).



\bibitem{butterfly_T3} J. Vidal, R. Mosseri, and B. Douçot,  Phys. Rev. Lett. \textbf{81},  5888 (1998).
\bibitem{DOS_Honeycomb} J. P. Hobson and W. A. Nierenberg,  Phys. Rev.  \textbf{89},  662 (1952).



\bibitem{butterfly} H. Aoki, M. Ando, and H. Matsumura, Phys. Rev. B {\bf 54}, 17296(R) (1996). 

\bibitem{butterfly_Honeycomb} R. Rammal, J. Physique \textbf{46},  1345 (1985).

\bibitem{butterfly_Kagome} T. Kimura, H. Tamura, K. Shiraishi, and H. Takayanagi,  Phys. Rev. B \textbf{65},  081307 (2002).





\bibitem{graphene_hall1} Y. Hatsugai, M. Kohmoto and Y.-S. Wu, Phys. Rev. B {\bf 54} 4898 (1996)
\bibitem{graphene_hall2} Y. Hatsugai, T. Fukui and H. Aoki, Phys. Rev. B {\bf 74}  205414 (2006)


\bibitem{graphene_hall}K. S. Novoselov, A. K. Geim, S. V. Morozov, D. Jiang, M. I. Katsnelson, I. V. Grigorieva, S. V. Dubonos, and A. A. Firsov, Nature {\bf 438}, 197 (2005); Y. Zhang, Y.-W. Tan, H. L. Stormer, and P. Kim, Nature {\bf 438}, 201 (2005).

\bibitem{gauge_field} J. Dalibard, F. Gerbier, G. Juzeli\=unas, and P. \"Ohberg, Rev. Mod. Phys. {\bf 83}, 1523 (2011).

\bibitem{lattice_gauge} M. Aidelsburger, M. Atala, S. Nascimb\`ene, S. Trotzky, Y.-A. Chen, and I. Bloch, Phys. Rev. Lett. {\bf 107}, 255301 (2011).

\bibitem{oktel} R. O. Umucalilar, H. Zhai and M. O. Oktel, Phys. Rev. Lett. {\bf 100}, 070402 (2008).

\bibitem{non_abelian_honeycomb} A. Bermudez, N. Goldman, A. Kubasiak, M. Lewenstein, and M. A. Martin-Delgado, New J. Phys. \textbf{12},  033041 (2010).
\bibitem{merging2} G. Montambaux, F. Pi�chon, J.-N. Fuchs, and M. O. Goerbig, Phys. Rev. B, {\bf 80} 153412 (2009).



\bibitem{decorated_honeycomb_TI} A. Ruegg, J. Wen and G. A. Fiete, Phys. Rev. B \textbf{81},  205115 (2010).

\bibitem{lieb_TI} C. Weeks and M. Franz,  Phys. Rev. B \textbf{82}, 085310 (2010).



\bibitem{na1} N. Goldman,I. Satija, P. Nikolic, A. Bermudez, M. A. Martin-Delgado, M. Lewenstein, and I. B. Spielman, Phys. Rev. Lett. {\bf 105} 255302 (2010).


\bibitem{na3} K. Osterloh, M. Baig, L. Santos, P. Zoller, and M. Lewenstein, Phys. Rev. Lett. {\bf 95} 010403 (2005).
\bibitem{na4} B. Beri and N. R. Cooper, Phys. Rev. Lett. {\bf 107}, 145301 (2011).


\bibitem{inversion1} L. Fu and C. L. Kane, Phys. Rev. B \textbf{76},  045302 (2007).
\bibitem{inversion2}T. L. Hughes, E. Prodan, and B. A. Bernevig,  Phys. Rev. B \textbf{83},  245132 (2011).

\bibitem{bragg2} X.-J. Liu, X. Liu, C. Wu and J. Sinova, Phys. Rev. A. {\bf 81}, 033622 (2010).
\bibitem{bragg} T. D. Stanescu, V. Galitski and S. Das Sarma, Phys. Rev. A {\bf 82}, 013608 (2010).
\bibitem{bragg3} N. Goldman, J. Beugnon and F. Gerbier, arXiv:1203.1246v1.



\bibitem{honeycomb_interaction} Z. Y. Meng, T. C. Lang, S. Wessel, F. F. Assaad, and  A. Muramatsu, Nature \textbf{464}, 847 (2010).
\bibitem{Lieb_interaction} W. Zhang, arXiv:1201.0722 (2012); W.-F. Tsai, C. Fang, H. Yao, and J. P. Hu, arXiv:1112.5789 (2011);
\bibitem{anyon} A. Kitaev, Ann. Phys.  {\bf 321}, 2  (2006).






\end{references}
\end{document}